\documentclass[12pt]{article}             
\begin{document}
\begin{center}
{\bf Photon--number tomography and fidelity}\\
O.~V.~Man'ko\\
 P.N.~Lebedev Physical Institute, Russian Academy of Sciences\\
      Leninskii Prospect, 53, Moscow 119991, Russia \\
      Email: omanko@sci.lebedev.ru
\end{center}
\begin{abstract}
The scheme of photon--number tomography is discussed in the
framework of star--product quantization. The connection of dual
quantization scheme and observables is reviewed. The quantizer and
dequantizer operators and kernels of star--product of tomograms in
photon--number tomography scheme and its dual one are presented in
explicit form. The fidelity and state purity are discussed in
photon--number tomographic scheme, and the expressions for
fidelity and purity are obtained in the form of integral of the
product of two photon--number tomograms with integral kernel which
is presented in explicit form. The properties of quantumness are
discussed in terms of inequalities on state photon--number
tomograms.
\end{abstract}
\vskip1cm

 \vspace{0.4cm}

\section*{Introduction}
The classical states are associated with the probability density.
The quantum states usually are associated either with the wave
function or the density operator. So, usually quantum and
classical states are described by different objects: probability
distribution function or operators. The probability representation
of quantum mechanics was suggested in \cite{Mancini96}, where the
quantum states are associated with tomographic probability
distribution called symplectic tomogram. The tomographic
probability representation of classical mechanics was suggested in
\cite{OlgaJRLR97},where the states of classical systems can be
also associated with tomograms. Thus, in the tomographic
probability representation, both classical and quantum states are
described by the same object -- tomogram. In
\cite{Dod97}
-\cite{andreev} the analogous description of quantum spin states
by probability distribution was suggested. The photon--number
tomography was introduced in \cite{vogel}-\cite{euroPLMancini}.
Photon--number tomograms of quantum gaussian states in one--mode
and multimode cases were obtained in \cite{mexica,JRLPNT}. The
symplectic and photon--number tomograms of photon--phonon mode in
the process of Raman Scattering were considered in
\cite{QuantumSemOpt}-
\cite{SPIEkuz}. The symplectic and photon--number tomograms of
even and odd coherent states were considered in
\cite{kotspie,izvran}. The symplectic tomograms of the states of
quantum resonant circuit and Josephson junction were considered in
\cite{Vax2011}. Explicit connection of photon--number tomogram
with measurable by homodyne detector optical tomogram
\cite{Bertran,Risken} was obtained in \cite{LaserPhysics,turku}.

The aim of the star--product approach is to find the description
of quantum properties by using classical--like instruments as
probability distributions. In the star--product quantization
scheme (see e.g. \cite{Stratonofich}
-\cite{Vasiliev}) functions on phase--space are used to describe
physical observables in quantum mechanics instead of operators.
The example of such approach is using the Wigner quasidistribution
function \cite{Wig32} instead of density operator for describing
quantum states. The Wigner function satisfies several conditions
which are satisfied by classical probability distributions on
phase--space, but the Wigner function can take the negative values
and can not be considered as the probability distribution. Due to
this it is called the quasiprobability.  Another star--product
scheme was suggested in \cite{MarmoOPhysScr}-
\cite{krakov} in which the quantum states can be described by the
standard positive probability distribution (symplectic tomogram)
\cite{Mancini96} instead of the description of the quantum states
by the Wigner function \cite{Wig32} which is quasidistribution.
The spin tomography was discussed in framework of star--product
quantization scheme in \cite{bradford}. The photon--number
tomography was considered in the framework of star--product
quantization in \cite{JRLPNT,Vax2008,Vax2010}. The connection of
dual quantization scheme and observables was found in
\cite{VitalePhysLet}.

The aim of the paper is to present a review of the photon--number
tomography within the framework of star--product quantization, to
discuss the connection of dual quantization scheme and
observables, to present in explicit form kernels of star--product
of symbols of operators in photon--number tomographic scheme and
its dual one, and to consider the fidelity and purity of the state
within the framework of the photon--number tomographic
representation.

The paper is organized as follows. In Sec. 1 we review the general
scheme of quantization based on star--product formalism. In Sec. 2
we review the dual quantization scheme and its connection with
observables. In Sec. 3 we review photon--number tomography
approach and present explicit expressions of kernels of
star--product of photon--number tomograms in initial and dual
schemes. In Sec. 4 the fidelity and purity of the state within the
framework of the photon--number tomographic representation are
considered. Conclusions are given in Sec. 5.

\section{General star--product quantization scheme}
Following \cite{MarmoOPhysScr,MarmoOJPA,krakov,VitalePhysLet}, let
us consider an operator $\hat A$ acting in a Hilbert space and the
$c$-number function $f_{\hat A}({\bf x})$ of vector variables
${\bf x}=(x_1,x_2,\ldots,x_n)$
\begin{equation}\label{SPeq.1}
f_{\hat A}({\bf x})=\mbox{Tr}\left[\hat A\hat{\cal U}({\bf
x})\right].
\end{equation}
We suppose that the relation (\ref{SPeq.1}) has the inverse
\begin{equation}\label{SPeq.2}
\hat A= \int f_{\hat A}({\bf x})\hat{\cal D}({\bf x})~d{\bf x}.
\end{equation}
The operator $\hat{\cal U}({\bf x})$ is called dequantizer
\cite{VitalePhysLet}. The function $f_{\hat A}({\bf x})$ is symbol
of the operator $\hat A$ and the operator $\hat{\cal D}({\bf x})$
is called quantizer \cite{VitalePhysLet}. The formulas
(\ref{SPeq.1}) and (\ref{SPeq.2}) are selfconsistent if one has
the following property of the quantizer and dequantizer
\begin{equation}\label{SVs15}
\mbox{Tr}\left[\hat{\cal U}({\bf x}) \hat{\cal D}({\bf
x}')\right]= \delta\left({\bf x}-{\bf x}'\right).
\end{equation}
We introduce the product of two functions corresponding to two
operators
\begin{eqnarray*}
\hat A\hat B \longrightarrow f_{\hat A}({\bf x})*f_{\hat B}({\bf
x}) \end{eqnarray*}
in the following form
\begin{eqnarray*}
f_{\hat A}({\bf x})* f_{\hat B} ({\bf x}):=\mbox{Tr}\left[\hat
A\hat B\hat{\cal U}({\bf x}) \right].
\end{eqnarray*}
The map (\ref{SPeq.1}) provides the nonlocal product of two
functions (star--product)
\begin{eqnarray*}
f_{\hat A}({\bf x})*f_{\hat B}({\bf x})=
\int f_{\hat A}({\bf x}'')f_{\hat B}({\bf x}') K({\bf x}'',{\bf
x}',{\bf x})\,d{\bf x}'\,d{\bf x}''.
\end{eqnarray*}
The kernel $K({\bf x}'',{\bf x}',{\bf x})$ of star--product of two
symbols is linear with respect to the dequantizer and nonlinear in
the quantizer operator
\begin{eqnarray}\label{kernel}
K({\bf x}'',{\bf x}',{\bf x})= \mbox{Tr}\left[\hat{\cal D}({\bf
x}'')\hat{\cal D}({\bf x}') \hat{\cal U}({\bf x})\right].
\end{eqnarray}
The standard product of operators is an associative product
$$ \hat A(\hat B \hat C)=(\hat A\hat B)\hat C.$$
The product of functions is associative since the product of the
operators is associative
\begin{eqnarray}\label{SPeq.6}
f_{\hat A}({\bf x})*\Big(f_{\hat B}({\bf x}) *f_{\hat C}({\bf
x})\Big)=
\Big(f_{\hat A}({\bf x})*f_{\hat B}({\bf x})\Big) *f_{\hat
C}({\bf x}).
\end{eqnarray}
The associativity condition for operator symbols means that the
kernel (\ref{kernel}) satisfies the nonlinear integral equation
\cite{VitalePhysLet}
\begin{eqnarray}\label{patC2}
\int K({\bf x}_1,{\bf x}_2, {\bf y})K({\bf y},{\bf x}_3, {\bf
x}_4)d{\bf y}=
\int K({\bf x}_1,{\bf y}, {\bf x}_4)K({\bf x}_2,{\bf x}_3, {\bf
y})d{\bf y}.
\end{eqnarray}

\section{The dual star--product scheme and quantum
observable}
Let us consider the following scheme
\begin{equation}\label{SPeq.1dual}
f^{(d)}_{\hat A}({\bf x})=\mbox{Tr}\left[\hat A\hat{\cal D}({\bf
x})\right],
\end{equation}
\begin{equation}\label{SPeq.2dual} \hat A= \int f^{(d)}_{\hat
A}({\bf x})\hat{\cal U}({\bf x})~d{\bf x}.
\end{equation}
One can see, that we permute the quantizer and the dequantizer
because the compatibility condition (\ref{SVs15}) will be valid in
both cases. We consider the new quantizer--dequantizer pair as
dual to the initial one
\begin{eqnarray*}
\hat{\cal U}^\prime({\bf x})\Longrightarrow \hat{\cal D}({\bf
x}),\quad \hat {\cal D}^\prime({\bf x})\Longrightarrow\hat{\cal
U}({\bf x}).
\end{eqnarray*}
The interchange corresponds to a specific symmetry of the equation
(\ref{patC2}) for associative star--product kernel. The
star--product of dual symbols~$f^{(d)}_{\hat A}({\bf x})$,
$f^{(d)}_{\hat B}({\bf x})$ of two operators $\hat A$ and $\hat B$
is described by dual integral kernel
\begin{eqnarray}\label{kerneldual}
K^{(d)}({\bf x}'',{\bf x}',{\bf x})=\mbox{Tr}\left[\hat{\cal
U}({\bf x}'')\hat{\cal U}({\bf x}')\hat{\cal D}({\bf x})\right].
\end{eqnarray}
The dual kernel (\ref{kerneldual}) is another solution of
nonlinear equation (\ref{patC2}) \cite{VitalePhysLet}.

Let us consider the mean value of quantum observable
$\hat A$
\begin{eqnarray*}
&&\langle \hat A\rangle=\mbox{Tr}\big(\hat\rho\hat A\big)= \int
f_{\hat \rho}({\bf x}) \mbox{Tr}\big(\hat{\cal D}(\bf x)\hat A
\big) d~\bf x\nonumber.
\end{eqnarray*}
Using the expression for dual symbol of operator
(\ref{SPeq.1dual}) we obtain the formula
\begin{eqnarray*}
\langle\hat A\rangle=
\int f_\rho(\bf x)f_{\hat A}^{(d)}(\bf x)\, d~\bf x.
\end{eqnarray*}
So, one can see, that the mean value of an observable $\hat A$ is
given by the integral of the product of the tomographic symbol of
the density operator $f_{\rho}(\bf x)$ in a given quantization
scheme and the symbol $f_{\hat A}^{(d)}(\bf x)$ of the observable
$\hat A$ in the dual scheme.

\section{Photon--number tomography as example of
star--product quantization}

Photon--number tomography is the method to reconstruct density
operator of quantum state using measurable probability
distribution function (photon statistics) called tomogram.
Photon--number tomography is different from optical tomography
method and from symplectic tomography scheme, where the continious
homodyne quadrature are measured for reconstructing quantum state.
In photon--number tomography the discrete random variable is
measured for reconstructing quantum state. The photon--number
tomogram
\begin{equation} \omega(n,\alpha)=\langle n\mid\hat
D(\alpha)\hat \rho\hat D^{-1}(\alpha)\mid n\rangle
\end{equation}
is the function of integer photon number $n$ and complex number
$\alpha$, $\hat\rho$ is the state density operator. The
photon--number tomogram is the photon distribution function (the
probability to have $n$ photons) in the state described by the
displaced density operator. We take Planck constant $\hbar=1$.

For example, photon--number tomogram for oscillator ground state
is
\[
w_0(n,\alpha)=\frac{e^{-|\alpha|^2}}{n!}|\alpha|^{2n}.
\]
The photon--number tomograms of excited oscillator states with
density operators $\hat\rho_m=\mid m\rangle\langle m \mid$ are
\[
w^{(m)}(n,\alpha)=\frac{n!}{m!}\mid\alpha\mid^{2(m-n)}e^{-\mid\alpha\mid^2}
\left(L_n^{m-n}(\mid\alpha\mid^2)\right)^2\]
\[m\geq n\,,\]
\[
w^{(m)}(n,\alpha)=\frac{m!}{n!}\mid\alpha\mid^{2(n-m)}e^{-\mid\alpha\mid^2}
\left(L_m^{n-m}(\mid\alpha\mid^2)\right)^2\]
\[m\leq n,\]
where $L_n^m(x)$ are Laguerre polynomials.

Let us consider photon--number tomogram in the framework of
star--product quantization. In the given photon--number tomography
quantization scheme the dequantizer is of the form
\begin{equation} \label{dequantizerphnumtom}
\widehat{\cal U}(\mbox{\bf x})=\hat D(\alpha)|n\rangle\langle
n|\hat D^{-1}(\alpha), \,\mbox{\bf x}=(n,\alpha).
\end{equation}
The quantizer operator is
\begin{equation}\label{quantizerphnumtom}
\widehat{\cal D}(\mbox{\bf x})=\frac{4}{\pi(1-s^2)}
\left(\frac{s-1}{s+1}\right)^{(\hat a^\dagger+\alpha^*)(\hat
a+\alpha)-n},
\end{equation}
where $s$ is ordering parameter, $\alpha$ is complex number
\[\alpha=\mbox{Re}\,\alpha+i\,\mbox{Im}\,\alpha,\]
$D(\alpha)$ is the Weyl displacement operator
\[\hat D(\alpha)=\exp(\alpha\hat a^{\dagger}-\alpha^*\hat a).\]
The kernel (\ref{kernel}) of star--product of photon--number
tomograms in the given photon--number tomography quantization
scheme is
\begin{equation}\label{kernelphotonnumbertommainscheme}
K(n_1,\alpha_1,n_2,\alpha_2,n_3,\alpha_3)=\mbox{Tr}\big[\hat{\cal
D}(n_1,\alpha_1)\hat{\cal D}(n_2,\alpha_2)\hat{\cal
U}(n_3,\alpha_3)\big].
\end{equation}
Putting in formula (\ref{kernelphotonnumbertommainscheme}) the
expressions for quantizer (\ref{quantizerphnumtom}) and
dequantizer (\ref{dequantizerphnumtom}) operators of
photon--number tomography scheme and taking the trace we obtain
the kernel of star--product of photon--number tomograms in the
explicit form
\begin{eqnarray}
&&
K(n_1,\alpha_1,n_2,\alpha_2,n_3,\alpha_3)=\Big(\frac{4}{\pi(1-s^2)}
\Big)^2\exp\large(it(n_1+n_2-2n_3)\large)
\exp\large[-|-\alpha_3+\alpha_1-\alpha_1e^{-it}\nonumber\\
&&\nonumber\\ && +\alpha_2
e^{-it}-\alpha_2e^{-2it}+\alpha_3e^{-2it}|^2 +\frac{1}{2}\large(
-\alpha_3\alpha_1^*+\alpha_3^*\alpha_1-\alpha_1\alpha_2^*
+\alpha_1^*\alpha_2 -\alpha_2\alpha_3^*+\alpha_2^*\alpha_3
+\alpha_3\alpha_1^*e^{it^*}\nonumber\\
&&\nonumber\\ && -|\alpha_1|^2e^{it^*}-\alpha_3\alpha_2^*e^{it^*}
+\alpha_1\alpha_2^*e^{it^*} -\alpha_3^*\alpha_1e^{-it}
 +|\alpha_1|^2e^{-it}
 +\alpha_3^*\alpha_2e^{-it}
-\alpha_1^*\alpha_2e^{-it}+\alpha_3\alpha_2^*e^{2it^*}\nonumber\\
&&\nonumber\\ && -\alpha_1\alpha_2^*e^{2it^*}
+\alpha_1\alpha_2^*e^{-it+2it^*}
-|\alpha_2|^2e^{-it+2it^*}-|\alpha_3|^2e^{2it^*}
+\alpha_1\alpha_3^*e^{2it^*}
-\alpha_1\alpha_3^*e^{-it+2it^*}\nonumber\\
&&\nonumber\\ && +\alpha_2\alpha_3^*e^{-it+2it^*}
-\alpha_3^*\alpha_2e^{-2it}
 +\alpha_1^*\alpha_2e^{-2it}
-\alpha_1^*\alpha_2e^{it^*-2it} -|\alpha_2|^2e^{it^*-2it}
+|\alpha_3|^2e^{-2it}-\alpha_1^*\alpha_3e^{-2it}\nonumber\\
&&\nonumber\\&& +\alpha_1^*\alpha_3e^{it^*-2it}
-\alpha_2^*\alpha_3e^{it^*-2it}\large)\large]
L_{n_3}\large(|-\alpha_3+\alpha_1-\alpha_1e^{-it}+\alpha_2
e^{-it}-\alpha_2e^{-2it}+\alpha_3e^{-2it}|^2\large),\label{kernephnum}
\end{eqnarray}
where
\[\frac{s-1}{s+1}=e^{it}\]
and $L_n(x)$ is Laguerre polynomial.

Let us consider the dual photon--number tomography quantization
scheme. We replace the quantizer (\ref{quantizerphnumtom}) and the
dequantizer (\ref{dequantizerphnumtom}) each the other and
consider the dual to the initial one quantizer--dequantizer pair.
So, the dequantizer operator in dual photon--number tomography
scheme is
\begin{equation}
\hat{\cal U}^\prime(n,\alpha)= \hat{\cal D}(n,\alpha),
\label{dequantizerphnumtomdual}
\end{equation}
the quantizer operator in dual photon--number tomography scheme
has the form
\begin{equation}\label{quantizerphnumtomdual}
\hat {\cal D}^\prime(n,\alpha)=\hat{\cal U}(n,\alpha).
\end{equation}
The kernel (\ref{kerneldual}) of star--product of symbols of
operators in the case of dual photon--number tomography
quantization scheme is
\begin{equation}\label{dualkernelphotonnumbertommainscheme}
K^{(d)}(n_1,\alpha_1,n_2,\alpha_2,n_3,\alpha_3)=\mbox{Tr}\big[\hat{\cal
U}(n_1,\alpha_1)\hat{\cal U}(n_2,\alpha_2)\hat{\cal
D}(n_3,\alpha_3)\big].
\end{equation}
Putting in formula (\ref{dualkernelphotonnumbertommainscheme}) the
expressions for quantizer (\ref{quantizerphnumtomdual}) and
dequantizer (\ref{dequantizerphnumtomdual}) operators of
photon--number tomography scheme and taking the trace we obtain
the dual kernel of star--product of symbols of operators (for
example, for the symbols of density operators -- photon--number
tomograms) in dual photon--number tomography scheme in the
explicit form
\begin{eqnarray}\label{kerneldualphnum}
&&\mbox{if}\quad n_1\geq n_2\,,\quad \mbox{then}\nonumber\\
&&\nonumber\\
&&K^{(d)}(n_1,\alpha_1,n_2,\alpha_2,n_3,\alpha_3)=\frac{4n_2!}{\pi(1-s^2)n_1!}
\exp\large(it(n_3-n_1)\large)\exp\large[\frac{1}{2}\large(-\alpha_1\alpha_2^*+\alpha_1^*\alpha_2\nonumber\\
&&\nonumber\\
&&-\alpha_2\alpha_3^*+\alpha_2^*\alpha_3
-\alpha_3\alpha_1^*+\alpha_3^*\alpha_1+\alpha_2\alpha_3^*e^{it^*}-|\alpha_3|^2e^{it^*}-\alpha_2\alpha_1^*e^{it^*}
+\alpha_3\alpha_1^*e^{it^*}-\alpha_3\alpha_2^*e^{-it}\nonumber\\
&&\nonumber\\
&&+|\alpha_3|^2e^{-it}+\alpha_1\alpha_2^*e^{-it}-\alpha_1\alpha_3^*e^{-it}
-|\alpha_2-\alpha_1|^2-|\alpha_3-\alpha_1-\alpha_3e^{-it}+\alpha_1e^{-it}|^2\large)\large]\nonumber\\
&&\nonumber\\
&&\times\large[\large(\alpha_2-\alpha_1\large)\large(-\alpha_3^*+\alpha_1^*+\alpha_3^*e^{it^*}
-\alpha_1^*e^{it^*}\large)\large]^{(n_1-n_2)}L_{n_2}^{n_1-n_2}\large(|\alpha_2-\alpha_1|^2\large)\nonumber\\
&&\nonumber\\
&&\times
L_{n_2}^{n_1-n_2}\large(|\alpha_3-\alpha_1-\alpha_3e^{-it}+\alpha_1e^{-it}|^2\large),\nonumber\\
&&\nonumber\\
&&\mbox{if}\geq n_2\rangle n_1\,,\quad \mbox{then}\nonumber\\
&&\nonumber\\
&&K^{(d)}(n_1,\alpha_1,n_2,\alpha_2,n_3,\alpha_3)=\frac{4n_1!}{\pi(1-s^2)n_2!}
\exp\large(it(n_3-n_1)\large)
\exp\large[\frac{1}{2}\large(-\alpha_1\alpha_2^*+\alpha_1^*\alpha_2\nonumber\\
&&-\alpha_2\alpha_3^*+\alpha_2^*\alpha_3
-\alpha_3\alpha_1^*+\alpha_3^*\alpha_1+\alpha_2\alpha_3^*e^{it^*}-|\alpha_3|^2e^{it^*}-\alpha_2\alpha_1^*e^{it^*}
+\alpha_3\alpha_1^*e^{it^*}-\alpha_3\alpha_2^*e^{-it}\nonumber\\
&&\nonumber\\&&+|\alpha_3|^2e^{-it}+\alpha_1\alpha_2^*e^{-it}-\alpha_1\alpha_3^*e^{-it}
-|\alpha_2-\alpha_1|^2-|\alpha_3-\alpha_1-\alpha_3e^{-it}+\alpha_1e^{-it}|^2\large)\large]
\nonumber\\
&&\nonumber\\
&&\times\large[\large(\alpha_2-\alpha_1\large)\large(-\alpha_3^*+\alpha_1^*+\alpha_3^*e^{it^*}
-\alpha_1^*e^{it^*}\large)\large]^{(n_1-n_2)}L_{n_1}^{n_2-n_1}\large(|\alpha_2-\alpha_1|^2\large)\nonumber\\
&&\nonumber\\
&&\times
L_{n_1}^{n_2-n_1}\large(|\alpha_3-\alpha_1-\alpha_3e^{-it}+\alpha_1e^{-it}|^2\large).
\end{eqnarray}
Both kernels (\ref{kernephnum}) and (\ref{kerneldualphnum}) are
solutions of equation (\ref{patC2}).

\section{Fidelity and purity in photon--number tomography scheme}
Now we calculate in terms of photon--number tomograms such
physical quantities as fidelity and purity. In fact, we have to
present the known quantities given in terms of density operators
in the form of integrals where the tomographic probability
distributions are involved. For continious photon quadratures the
generalized fidelity which equals to trace of product of $n$
density operators was calculated by using expression of the
density operators in terms of the symplectic tomograms in
\cite{fortsch}. In the framework of photon--number tomography
scheme we get for the fidelity the expression
\begin{equation}\label{fidelity} {\cal F}=
\mbox{Tr}(\hat\rho_1\hat\rho_2)=\sum_{n_1,n_2=0}^{\infty}\int
w_1(n_1,\alpha_1)w_2(n_2,\alpha_2){\cal
K}(n_1,n_2,\alpha_1,\alpha_2)\,d\,\alpha_1\,d\,\alpha_2
\end{equation}
and for the state purity the analogous expression
\begin{equation}\label{purity}
{\cal P}=
\mbox{Tr}(\hat\rho^2)=\sum_{n_1,n_2=0}^{\infty}\int
w(n_1,\alpha_1)w(n_2,\alpha_2){\cal
K}(n_1,n_2,\alpha_1,\alpha_2)\,d\,\alpha_1\,d\,\alpha_2,
\end{equation}
where the kernel ${\cal K}(n_1,n_2,\alpha_1,\alpha_2)$ is of the
form
\begin{eqnarray}\label{kernelfidelityphnum}
&&{\cal K}(n_1,n_2,\alpha_1,\alpha_2)=
e^{it(n_1+n_2)}\exp\large[\frac{1}{2}\large(-|\alpha_1|^2e^
{it^*}+|\alpha_1|^2e^ {-it}+\alpha_1\alpha_2^*(1-e^{-it})(e^
{it^*}-e^ {-2it^*})\nonumber\\
&&\nonumber\\
&&-\alpha_1^*\alpha_2(1-e^{it^*})(e^{-it}-e^{-2it})\large)-|\alpha_1-\alpha_1
e^{-it}+\alpha_2e^{-it}-\alpha_2e^{-2it}|^2-e^{-2it}\large]\nonumber\\
&&\nonumber\\
&&\times{\cal J}_0\large(2e^{-it}|\alpha_1-\alpha_1
e^{-it}+\alpha_2 e^{-it}-\alpha_2 e^{-2it}|\large),
\end{eqnarray}
and ${\cal J}_0(x) $ is Bessel function. So, we obtain the
expressions for fidelity (\ref{fidelity}) and purity
(\ref{purity}) in the form of integral of the product of two
photon--number tomograms with integral kernel ${\cal
K}(n_1,n_2,\alpha_1,\alpha_2)$, which we obtain in explicit form
(\ref{kernelfidelityphnum}).

We have following inequalities for fidelity (\ref{fidelity}) and
purity (\ref{purity}) of quantum states of real physical system
written in terms of photon--number tomograms
\begin{equation}\label{fidelityphnumtom}
0\leq\sum_{n_1,n_2=0}^{\infty}\int
w_1(n_1,\alpha_1)w_2(n_2,\alpha_2){\cal
K}(n_1,n_2,\alpha_1,\alpha_2)\,d\,\alpha_1\,d\,\alpha_2 \leq
1,\end{equation} \begin{equation}\label{purityphnumtom}
0\leq\sum_{n_1,n_2=0}^{\infty}\int
w(n_1,\alpha_1)w(n_2,\alpha_2){\cal
K}(n_1,n_2,\alpha_1,\alpha_2)\,d\,\alpha_1\,d\,\alpha_2, \leq
1.
\end{equation}
Also photon--number tomograms associated with quantum states of
real physical system must satisfy inequalities which are
conditions of density operators nonnegativity
\begin{equation}\label{nonnegativityphnumtom}
\frac{4}{\pi(1-s^2)}\sum_{n=0}^\infty \int
\left(\frac{s-1}{s+1}\right)^{(\hat a^\dagger+\alpha^*)(\hat
a+\alpha)-n} w(n,\alpha)\,d\,\alpha\geq 0.
\end{equation}
For the photon states we obtained expressions for fidelities
(\ref{fidelityphnumtom}) and purities (\ref{purityphnumtom}) given
in terms of photon--number tomograms which are probability
distributions. The expressions (\ref{fidelityphnumtom}),
(\ref{purityphnumtom}) and (\ref{nonnegativityphnumtom}) can be
used for checking quantumness of the states analogous to
\cite{Bellini}, where the expressions written in terms of
measurable optical tomograms \cite{fortsch} were used for checking
quantumness of the states in experiments with homodyne detection.
The inequality (\ref{nonnegativityphnumtom}) can be violated for
the classical electromagnetic field.

\section*{Conclusion}
We review the notion of quantum state in photon--number tomography
approache. The scheme of photon--number tomography is discussed in
the framework of star--product quantization. As new results
presented in the paper we want to mention the explicit expressions
of the kernels of star--product of photon--number tomograms:
expression (\ref{kernephnum}) in given and expression
(\ref{kerneldualphnum}) in dual quantization schemes. The fidelity
and state purity are discussed in the framework of the
photon--number tomography scheme and the explicit expressions for
them in the form of the product of two photon--number tomograms
with integral kernel (\ref{kernelfidelityphnum}), which is
obtained in explicit form, are presented. The properties of
quantumness and classicality are discussed in terms of
inequalities on state photon--number tomograms.

\subsection*{Acknowledgments}
The study was supported by the Russian Foundation for Basic
Research under Project No.~10-02-00312. The author thanks the
Organizers of the Conference Quantum Theory Reconsideration of
Foundations - 6 and especially Prof. A. Khrennikov for invitation
and kind hospitality.

\end{document}